\documentclass[12pt,preprint]{aastex}

\usepackage{amsmath}

\begin{document}

\title{A Population of Ultraluminous X-ray Sources with An Accreting
Neutron Star }
\author{
Yong Shao$^{1,2}$ and Xiang-Dong Li$^{1,2}$}

\affil{$^{1}$Department of Astronomy, Nanjing University, Nanjing
210046, China; lixd@nju.edu.cn}

\affil{$^{2}$Key laboratory of Modern Astronomy and Astrophysics (Nanjing University),
Ministry of Education, Nanjing 210046, China}

%\affil{$^{}$}

\begin{abstract}
Most ultraluminous X-ray sources (ULXs) are believed to be X-ray
binary systems, but previous observational and theoretical studies
tend to prefer  a black hole rather than a neutron star accretor.
The recent discovery of 1.37 s pulsations from the ULX M82 X-2 has
established its nature as a magnetized neutron star. In this work we
model the formation history of neutron star ULXs in an M82- or Milky
Way-like galaxy, by use of both binary population synthesis and
detailed binary evolution calculations. We find that the birthrate
is around $10^{-4}\, \rm yr^{-1}$ for the incipient X-ray binaries
in both cases. We demonstrate the distribution of the ULX population
in the donor mass - orbital period plane. Our results suggest that,
compared with black hole X-ray binaries, neutron star X-ray binaries
may significantly contribute to the ULX population, and high-mass
and intermediate-mass X-ray binaries dominate the neutron star ULX
population in M82- and Milky Way-like galaxies, respectively.

\end{abstract}


\section{Introduction}

Ultraluminous X-ray sources (ULXs) are off-nuclear, point-like
sources with X-ray luminosities exceeding $ 10^{39}$ ~ergs$^{-1}$,
first discovered in nearby galaxies with \textit{Einstein}
\citep{f89}. More recent observations with improved X-ray
telescopes, such as \textit{Chandra} and \textit{XMM-Newton}, have
greatly increased the number of this kind of sources \citep[][for
reviews]{f06,r07,f11}.
%discovered in nearby galaxies roughly corresponding to the
%Eddington luminosity for a stellar-mass black hole
They are most likely  X-ray binaries (XRBs), in which a compact
object accretes from a donor star through Roche-lobe overflow
(RLOF), but the nature of these objects has not been completely
uncovered. If the radiation is isotropic and below the Eddington
limit, the extremely high luminosities imply the presence of an
accreting intermediate-mass ($10^2-10^5\,M_{\sun}$) black hole (BH)
\citep{cm99}. Alternatively they are believed to be stellar-mass BHs
with super-Eddington accretion. For example, \citet{k01} proposed
that geometrical beaming in the case of rapid accretion could lead
to a very high apparent luminosity for a stellar-mass BH.
\citet{b02} showed that the isotropic luminosity of an accreting BH
can exceed the Eddington limit by a factor of about 10 due to the
photon-bubble instability in the accretion disk. Recently, the
masses of two ULXs were dynamically measured to be in the
stellar-mass BH range \citep{liu13,m14}.

Since the X-ray luminosities of ULXs are significantly higher than
the Eddington limit $L_{\rm E}$ (around $2\times10^{38} \rm erg$
$\rm s^{-1} $)  for a $1.4M_{\odot} $ neutron star (NS)\footnote{If
the NS possesses a strong magnetic field to channel the accreting
material onto its surface, the critical luminosity could be higher,
about $\frac{l_0}{2\pi d_0}L_{\rm E}\sim 4(\frac{l_0/d_0}{25})L_{\rm
E}$,  where $l_0$ and $d_0$ are the length and the thickness of the
accreting funnel, respectively \citep{b76}.}, it is challenging to
explain these bright X-ray sources with an accreting NS. However,
the discovery of 1.37 s pulsations from the ULX M82 X-2 has provided
unambiguous evidence for its NS nature \citep{b14}, implying that
accreting NSs also contribute to the ULX population.

M82 X-2 is in a 2.53 day orbit with a companion star more massive
than $5.2~M_{\sun}$ \citep{b14}. The extremely high (isotropic)
luminosity (around $10^{40}$ ergs$^{-1}$) \citep{f10} and rapid
spin-up (at a rate of $-2\times 10^{-10}$ ss$^{-1}$) \citep{b14}
clearly indicate that the binary is undergoing rapid mass transfer.
Since the companion star is significantly more massive than the NS,
the mass transfer is subject to delayed dynamical instability, and
must be currently in the early phase of RLOF. After that the NS will
be engulfed by the transferred matter, resulting in the formation of
a common envelope (CE) \citep{b91}. The objective of this paper is
to investigate how many such accreting NSs can be responsible for
ULXs in a galaxy like M82 or the Milky Way (MW). We first explore
the properties of the incipient NS XRBs, using a binary population
synthesis (BPS) method (in Sect.~2), then calculate the detailed
evolutions of these XRBs to obtain the numbers and the luminosity
functions of the ULXs (in Sect.~3). We summarize our results in
Sect.~4.

\section{Generation of the incipient NS XRBs}

To model the formation history of NS XRB-ULXs, we adopt the BPS code
initially developed by \citet{h02} to calculate the evolution of a
large population of the primordial binaries. We have updated and
modified the code in several aspects \citep[see][for details]{sl14},
especially the conditions for dynamically stable mass transfer and
the treatments of CE evolution, which are briefly described as
follows.

During the evolution of a primordial binary, the primary first
evolves off the main sequence and expands in size. This can lead to
RLOF onto the secondary, causing it to be spun up and rejuvenated.
If the mass transfer proceeds so slowly that the secondary can
remain in thermal equilibrium, the mass transfer is thought to be
stable. Otherwise the secondary will get out of thermal equilibrium
and expand. This expansion may finally cause the secondary to fill
its own RL, leading to the formation of a contact binary
\citep{ne01}. A critical mass ratio $q_{\rm cr}$ is usually used to
determine whether or not the mass transfer is dynamically stable in
a binary. Instead of using the empirical results for $q_{\rm cr}$ of
\citet{h02}, \citet{sl14} numerically calculate it considering both
the response of the secondary to mass accretion and the effect of
possible mass loss. Here we use the results of Model II given by
\citet{sl14}, in which it is assumed that half of the transferred
mass is accreted by the secondary, and the other half is lost from
the system. This model can well fit the observational distribution
of Be/X-ray binaries.

When the mass transfer is dynamically unstable, we employ the
standard energy conservation equation \citep{w84} to deal with the
subsequent CE evolution, that is, in the spiral-in phase, the
orbital energy of the secondary is used to expel the envelope of the
primary. When calculating the binding energy of the stellar
envelope, we include the contribution from the internal energy and
adopt the fitting formulae of \citet{xl10} for the binding energy
parameter $\lambda$. We assume the CE efficiency $ \alpha_{\rm CE}=
1.0$ in our calculations.

The evolution of a binary is determined by the primary mass $ M_{1}
$, secondary mass $ M_{2} $, and orbital angular momentum
\citep{h02}. The orbit is invariably circularized before interaction
by standard tidal interactions, so all binary orbits are taken to be
circular. We adopt the initial mass function given by \citet{k93}
for the primary stars,
%\begin{equation}
%\Phi(M)\propto M^{-2.7},
%\Phi(M) =\left\{
%    \begin{array}{ccc}
%      a_{1}M^{-1.3}  & \mbox{if} & 0.08 \leq M < 0.5,  \\
%      a_{2}M^{-2.2}  & \mbox{if} & 0.5 \leq  M <0.2,   \\
%      a_{2}M^{-2.7}  & \mbox{if} & 1.0 \leq  M < \infty,
 %   \end{array}  \right.
%\end{equation}
%
and a flat distribution between 0 and 1 for the initial mass ratio
of the secondary to the primary.
%In our calculation we choose the
%primary mass to distribute in the range of $ 7-40 M_{\odot} $
%\citep[relevant to NS formation,][]{b08}, and the secondary mass
%between $ 1M_{\odot}$ and $30 M_{\odot}$, to guarantee that almost
%all NS XRBs can be included.
For the initial orbital separation $a$, we assume that $\ln a$ is
evenly distributed between $a=3R_{\odot}$ and $10^{4}R_{\odot}$. We
adopt solar metallicity $Z = 0.02$ in our simulations.

We consider two cases of star formation activities for star forming
galaxies like M82 and late-type galaxies like the MW. We adopt a
constant overall star formation rate of $10 M_{\odot}\, {\rm
yr}^{-1} $ over the last 100 Myr \citep{f03,y09} in case (1) , and a
constant star formation rate $3 M_{\odot}\, {\rm yr}^{-1} $ over the
13 Gyr period in case (2).
%\textbf{We follow the
%method of \citet{h02} to deal with the birthrate of a particular
%population.}

%\textbf{Form the evolution of primordial binaries, the incipient
%XRBs can be produced. The envelope of the primary may be striped
%through the processes of either dynamically stable mass transfer or
%common envelope evolution, the remnant of the primary may eventually
%experience a supernova explosion to form an NS. The newborn NS has a
%kick velocity that may lead to the disruption of the binary system.
%So that only a small amount of primordial binaries can evolve into
%the incipient XRBs, surviving from the NS kick.}

We have evolved a population of $ 7\times10^{6} $ primordial
binaries, and generated a subset of about $ 1.0\times10^{4}$ and $
1.5\times10^{4}$ incipient XRBs containing a NS and an unevolved
secondary star of mass lower than $20 M_{\odot} $ in cases (1) and
(2), respectively. Note that for the mechanisms of NS formation we
consider both core-collapse supernovae of massive stars and
electron-capture supernovae of intermediate-mass stars, following
the criterion suggested by \citet{f12}. In Fig.~1 we plot the
distribution of these XRBs in the donor (or secondary) mass $(M_{\rm
d})$ - orbital period ($P_{\rm orb}$) plane (the left panel) and the
birthrate distribution as a function of the donor mass and of the
orbital period (right panel). Note that in case (1) most of the
secondaries are more massive than $5\,M_{\sun}$ (upper panel),
because of the much shorter evolution time than in case (2) (lower
panel). The total birthrates are estimated to be about
$1.6\times10^{-4} $ $ \rm yr^{-1}$ and $6.9\times10^{-5} $ $ \rm
yr^{-1}$, respectively.

\section{Evolution to ULXs}

Based on the result in Fig.~1 we calculate the evolution of the
generated NS XRBs with the TWIN version of the stellar evolution
code developed by \citet{e71,e72}. Here the initial mass of the NS
is assumed to be $ 1.4M_{\odot}$. We have evolved thousands of
binary systems with the donor mass $ M_{\rm d} $ varying from $
0.5M_{\odot} $ to $ 20M_{\odot} $ by steps of $0.25 M_{\odot}$, and
the orbital period $ P_{\rm orb} $ (in units of days) increasing
logarithmically from $-$0.5 to 3 by steps of 0.1, as obtained from
Fig.~1. We use these binaries to represent all the XRBs, and their
numbers are calculated by cumulating the XRBs in a specific matrix
of $\Delta M_{\rm d}\times \Delta (\log P_{\rm orb}) $  in the $
M_{\rm d}-P_{\rm orb} $ plane by weighing their formation rates and
life spans. Note that the XRBs are initially eccentric due to the
supernova kicks, which are assumed to have a Maxwellian distribution
with a dispersion $\sigma=265$ kms$^{-1}$ \citep{h05} for
core-collapse supernovae and 50 kms$^{-1}$ \citep{d06} for
electron-capture supernovae. Here we assume that the orbital angular
momentum of an incipient XRB is conserved and then the system is
quickly circularized with a new separation, which is smaller by a
factor of $(1-e^{2})$ \citep{b91}.

In Eggleton's code mass transfer via RLOF can be modeled as a
function of the potential difference $\Delta\phi$ between the
stellar surface and the RL surface. When one of the stars overfills
its RL, the mass flux at each mesh point outside the Roche surface
is given by
\begin{equation}
\frac{d\dot{M}}{dm}=-10^4\frac{\sqrt{2\Delta\phi}}{r},
\end{equation}
where $m$ and $r$ are the mass coordinate and the radius,
respectively. The mass transfer rate is then calculated by
integrating Eq. ~(1) over all mesh points outside the Roche surface
potential.  Actually before the photospheric radius of a massive
donor star reaches its RL, the atmospheric matter begins to spill
over towards the NS along the inner Langrangian point. This phase of
beginning atmospheric RLOF precedes the main phase of RLOF until the
mass transfer rate rises to the Eddington value \citep{s79}.
Subsequently the mass transfer rate increases rapidly to become
super-Eddington. In Fig.~2 we show the exampled evolutionary tracks
of three binary systems. The initial parameters are $ M_{\rm
d}=6M_{\odot}$ and $ P_{\rm orb}  = 1$  d, $ M_{\rm d}=6M_{\odot}$
and $ P_{\rm orb}  = 10$  d, and $ M_{\rm d}=10 M_{\odot}$ and $
P_{\rm orb} = 1$  d in the top, middle, and bottom panels,
respectively. In the top panel, the donor evolves to overflow from
the RL at the age of $ 39.4$ Myr. The mass transfer rate  increases
from $2.1\times10^{-10} M_{\odot}\, {\rm yr}^{-1}$ to $
2.4\times10^{-4}
 M_{\odot}\, {\rm yr}^{-1}$ within a time of $ 4.46\times10^{5} $ yr.
The orbital period decays to 0.58 d, and the donor mass decreases to
$ 5.59 M_{\odot} $, which means that $ 0.41 M_{\odot} $ of the
donor's envelope is stripped before the CE occurrence. In the middle
panel, a  longer initial orbital period of 10 d is set for the
binary. The onset of RLOF occurs at the age of 66.08 Myr, and the
donor star is more evolved. The mass transfer rate rises from $
1.3\times 10^{-10}M_{\odot}\, {\rm yr}^{-1}$ to $1.1\times10^{-3}
M_{\odot}\, {\rm yr}^{-1}$ within a shorter time of $
1.2\times10^{4} $ yr. The orbital period drops to $ 3.68 $ d, and $
0.78 M_{\odot} $ material is transferred from the donor within this
time. In the bottom panel, the initial donor star is more massive
(with a mass of $ 10M_{\odot} $), and the mass transfer lasts $
2.32\times10^{5} $ yr. The orbital period decreases to 0.75 d when
0.17 $ M_{\odot} $ of the donor's envelope is transferred.

Given the mass transfer rate, we calculate the X-ray luminosity in
two ways. In the first one, we calculate it with the traditional
formula
\begin{equation}
L_{\rm X}=0.1\dot{M}c^2,
\end{equation}
without considering the Eddington limit. In the second, we use the
same formula for sub-Eddington accretion rates. When $\dot{M}$ is
higher than the Eddington accretion rate $ \dot{M}_{\rm E}$, we
adopt the model of \citet{k08} to convert the mass transfer rates
into the X-ray luminosities. In this model, the accretion disk
becomes geometrically thick, which influences the X-ray luminosity
in two ways. First, radiation becomes less efficient and  the
bolometric luminosity no longer follows $\dot{M}$ linearly. Second,
the outgoing radiation may be collimated due to a biconical geometry
at the inner part of the accretion disk. The accretion luminosity is
then contributed by two parts. The region outside the so-called
spherization radius $ R_{\rm sph} $ where the mass inflow first
becomes locally Eddington \citep{ss73,b06} releases the accretion
luminosity close to $L_{E}$. For the region within $ R_{\rm sph}$,
the accretion luminosity is about $\ln(R_{\rm sph}/3R_{\rm S})\sim
\ln(\dot{M}/\dot{M}_{\rm E})$, where $R_{\rm S}$ is the
Schwarzschild radius \citep{f02}. The total luminosity is then
\citep{k08},
\begin{equation}
L_{\rm acc} \simeq L_{E}\left[1+\ln \left(
\frac{\dot{M}}{\dot{M}_{E}} \right) \right].
\end{equation}
%where $ \dot{M}_{\rm E} $ and $ \dot{M}_{\rm d} $ are  and the mass accretion rate, respectively.
Because of the geometric collimation, one can see the source in
directions within one of the cones, with an apparent (isotropic)
X-ray luminosity
\begin{equation}
L_{\rm X} \simeq \frac{L_{E}}{b}\left[1+\ln \left(
\frac{\dot{M}}{\dot{M}_{E}} \right) \right],
\end{equation}
where $ b$ is the beaming factor, possibly depending on $\dot{M}$.
Here we adopt $ b\sim0.1 $ as suggested by \citet{k08}.

In Fig.~3 we present the number distribution of the predicted ULX
population in case (1) (i.e., in a galaxy like M82) as a function of
$ M_{\rm d}$ and $P_{\rm orb}$,  with $L_{\rm X}$ greater than
$10^{39}$~ergs$^{-1} $. In the upper and lower panels the X-ray
luminosities are calculated with Eqs. (2) and (4), respectively. As
noted before, for each matrix element in the left panel, the number
is calculated by multiplying the birthrate of the incipient XRBs
with the evolutionary time span within the matrix element. It is
seen that most ULXs tend to be high-mass XRBs in relatively short
orbits (with orbital periods shorter than a few days). Figure 4
shows the same distributions in case (2) (i.e., in a MW-like
galaxy). A comparison of Figs.~3 and 4 shows that in the latter case
XRBs with donor of mass lower than $3M_{\odot}$ dominate the
population because of their much longer lifetime.

In Fig.~5 we plot the X-ray luminosity function of the ULX
population in cases (1) (left) and (2) (right).  The red and black
curves correspond to $ M_{\rm d}$ higher than $2M_{\odot} $ and $
5M_{\odot} $, respectively. For the dashed and solid curves the
luminosities are calculated with Eqs.~(2) and (4), respectively. In
case (2) the effect of anisotropic radiation on the observable
number is taken into account. The predicted ULX numbers lie between
a few tenths and a few in each case. \citet{m08} investigated the
evolution of ULXs consisting of a stellar-mass BH accretor, and
found that their numbers range from 0.1 to 0.2 for $L_{\rm X}$
higher than $2\times 10^{39}$ ergs$^{-1}$ in a MW-like galaxy. Hence
accreting NSs may play an important role in the formation of ULXs,
compared with BHs.

\section{Summary}
We have calculated the formation history of NS ULXs. Our results
show that, (1) compared with BH XRBs, NS XRBs may significantly
contribute to the ULX population; (2) high-mass and
intermediate-mass XRBs dominate the ULX population in M82- and
MW-like galaxies, respectively.

\begin{acknowledgements}

This work was supported by the Natural Science Foundation of China
under grant numbers 11133001, 11203009, and 11333004, the Strategic
Priority Research Program of CAS (under grant number XDB09000000),
and the graduate innovative project of Jiangsu Province (CXZZ13-0043).

\end{acknowledgements}

\clearpage

\begin{figure}

\includegraphics[scale=0.5]{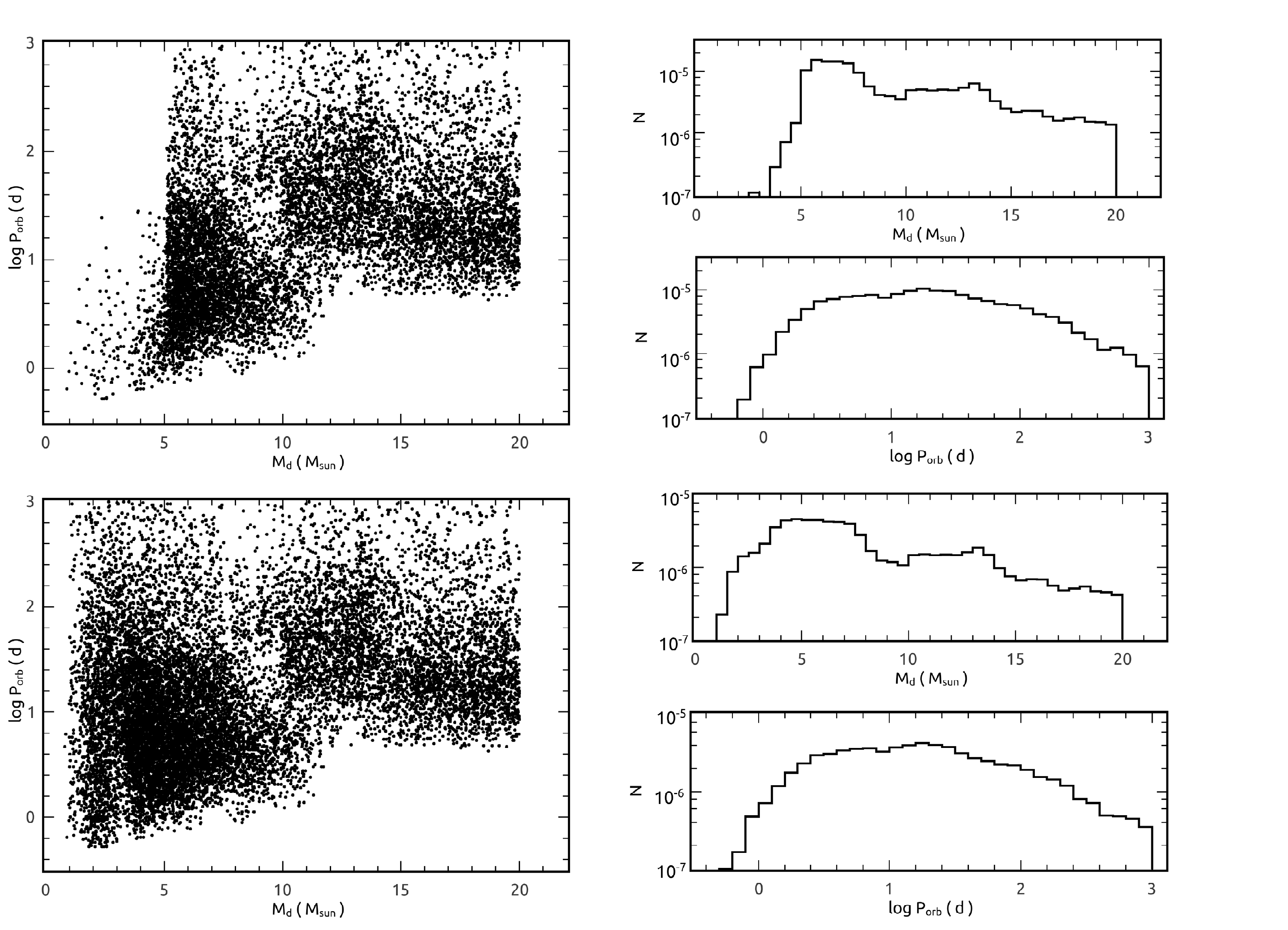}

\caption{The upper and lower panels show the predicted distributions
of the incipient NS XRBs and their birthrates in cases (1) and (2),
respectively. The left panel shows the XRB distribution in the donor
mass vs. orbital period plane. Their birthrate distributions are
presented in the right panel. \label{figure1}}

\end{figure}

\clearpage

\begin{figure}[h,t]

\includegraphics[scale=0.6]{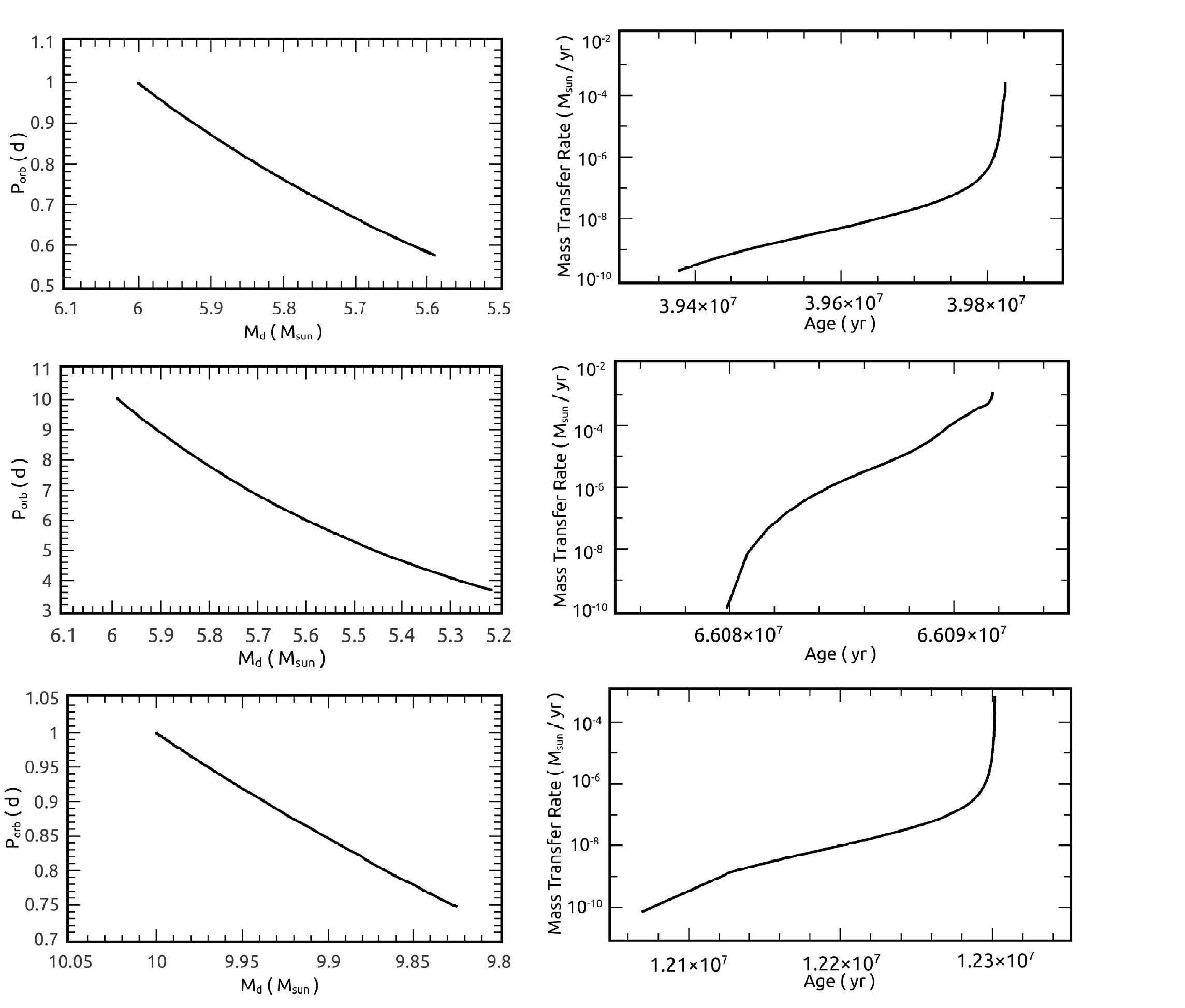}

\caption{Example evolution of the orbital period and the mass
transfer rate for three NS XRBs, as a function of the donor mass and
the age, respectively. The initial parameters are $ M_{\rm
d}=6M_{\odot}$ and $ P_{\rm orb}  = 1$  d, $ M_{\rm d}=6M_{\odot}$
and $ P_{\rm orb}  = 10$  d, and $ M_{\rm d}=10 M_{\odot}$ and $
P_{\rm orb} = 1$  d in the top, middle, and bottom panels,
respectively.
  \label{figure2}}

\end{figure}

\clearpage

\begin{figure}[h,t]

\includegraphics[scale=0.6]{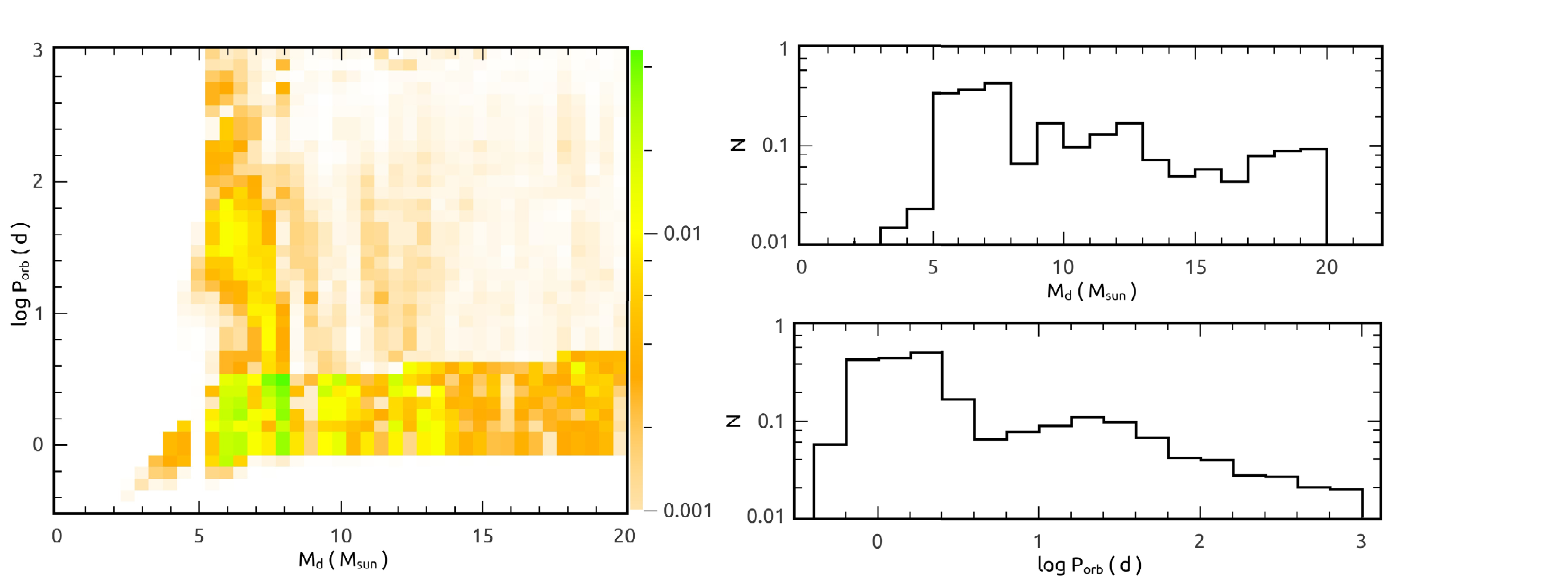}
\includegraphics[scale=0.6]{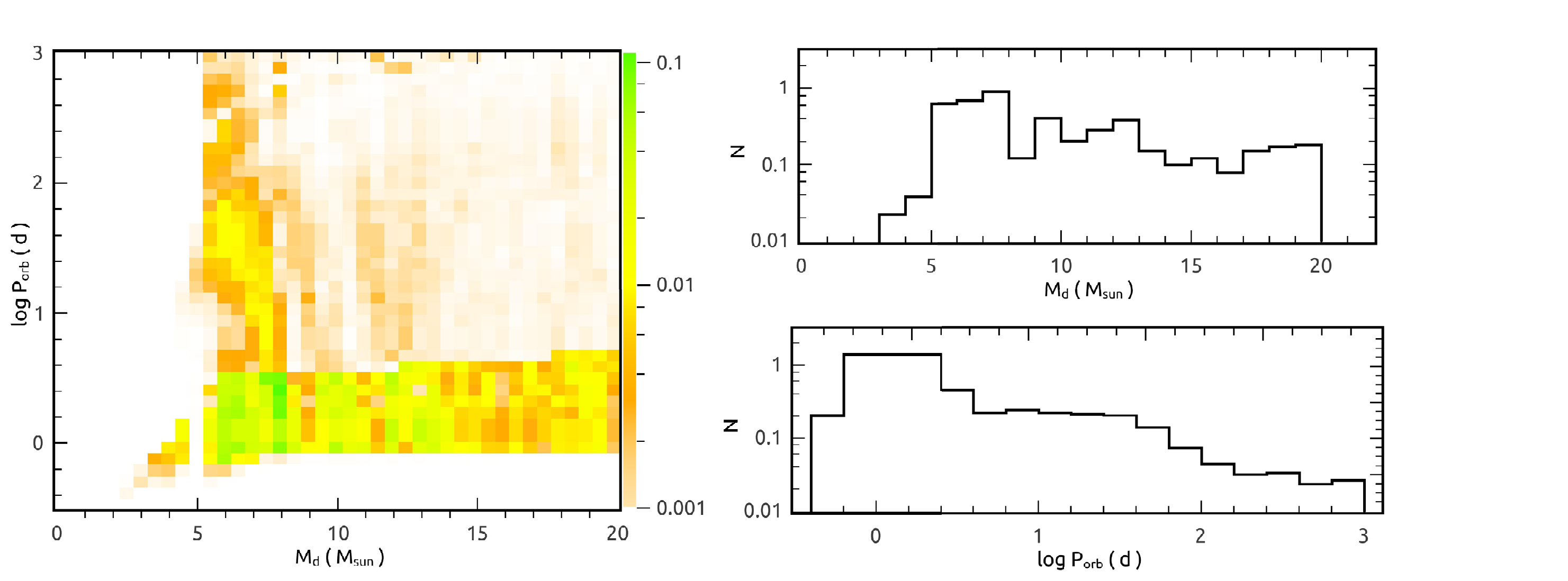}

\caption{The predicted distributions of NS ULXs in case (1). In the
upper and lower panels, the X-ray luminosities are calculated with
Eqs.~(2) and (4), respectively. The left panel shows the ULX
distribution in the $ M_{\rm d}-P_{\rm orb} $ plane. The color in
each matrix element represents the number of ULXs with $L_{X}$
higher than $10^{39}$ ergs$^{-1} $. The right panel shows the number
distribution of the ULXs as a function of the donor mass $ M_{\rm
d}$ and the orbital period $P_{\rm orb} $. \label{figure3}}

\end{figure}

\clearpage

\begin{figure}[h,t]

\includegraphics[scale=0.6]{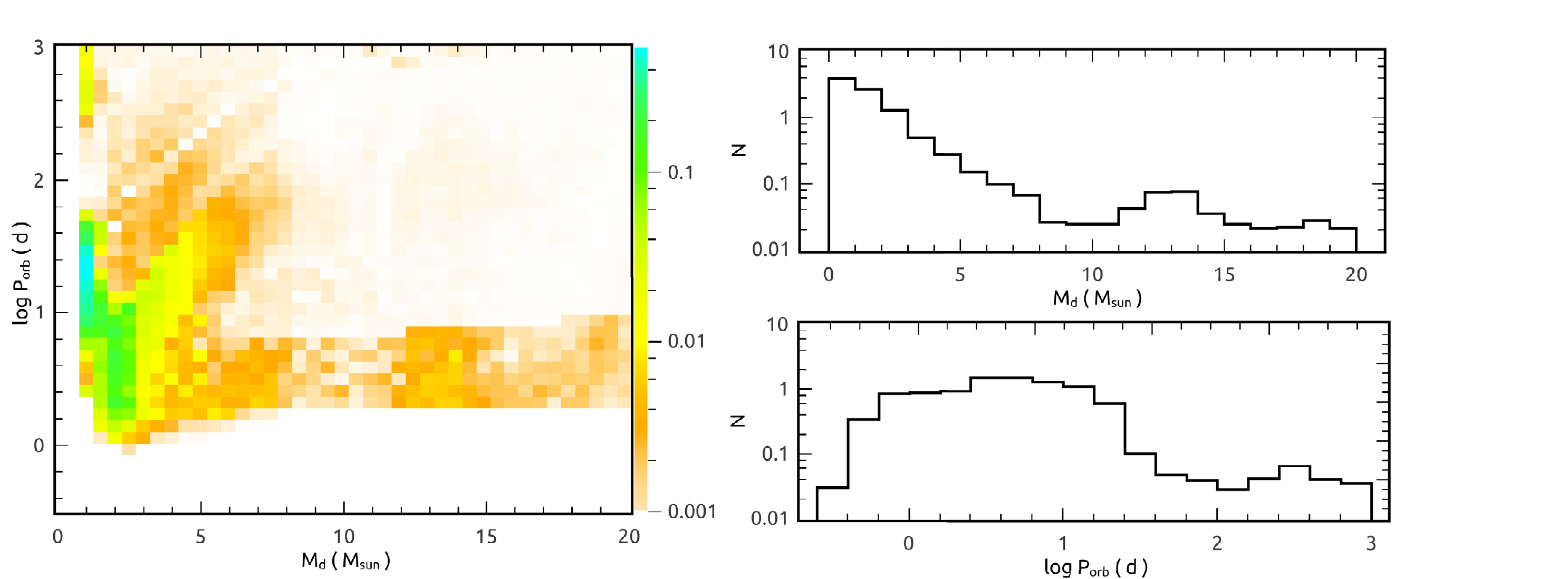}
\includegraphics[scale=0.6]{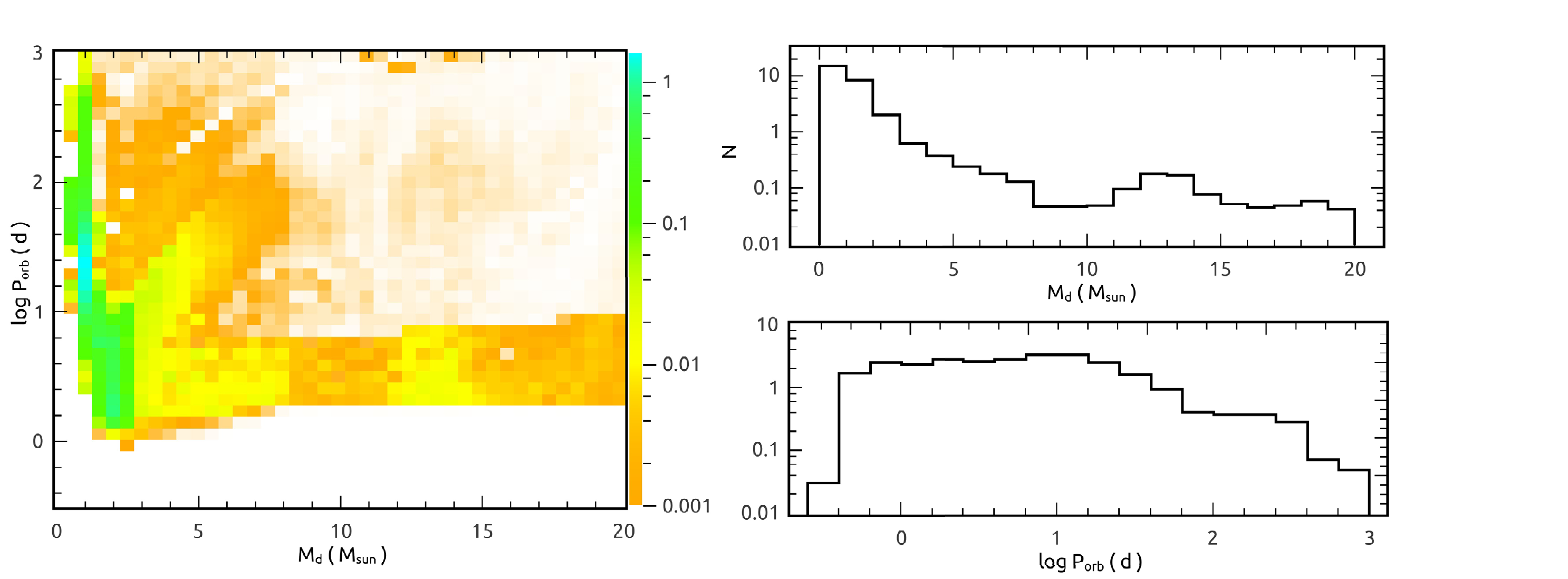}

\caption{Same as Fig.~3, but for case (2). \label{figure4}}

\end{figure}

\clearpage

\begin{figure}[h,t]

\plottwo{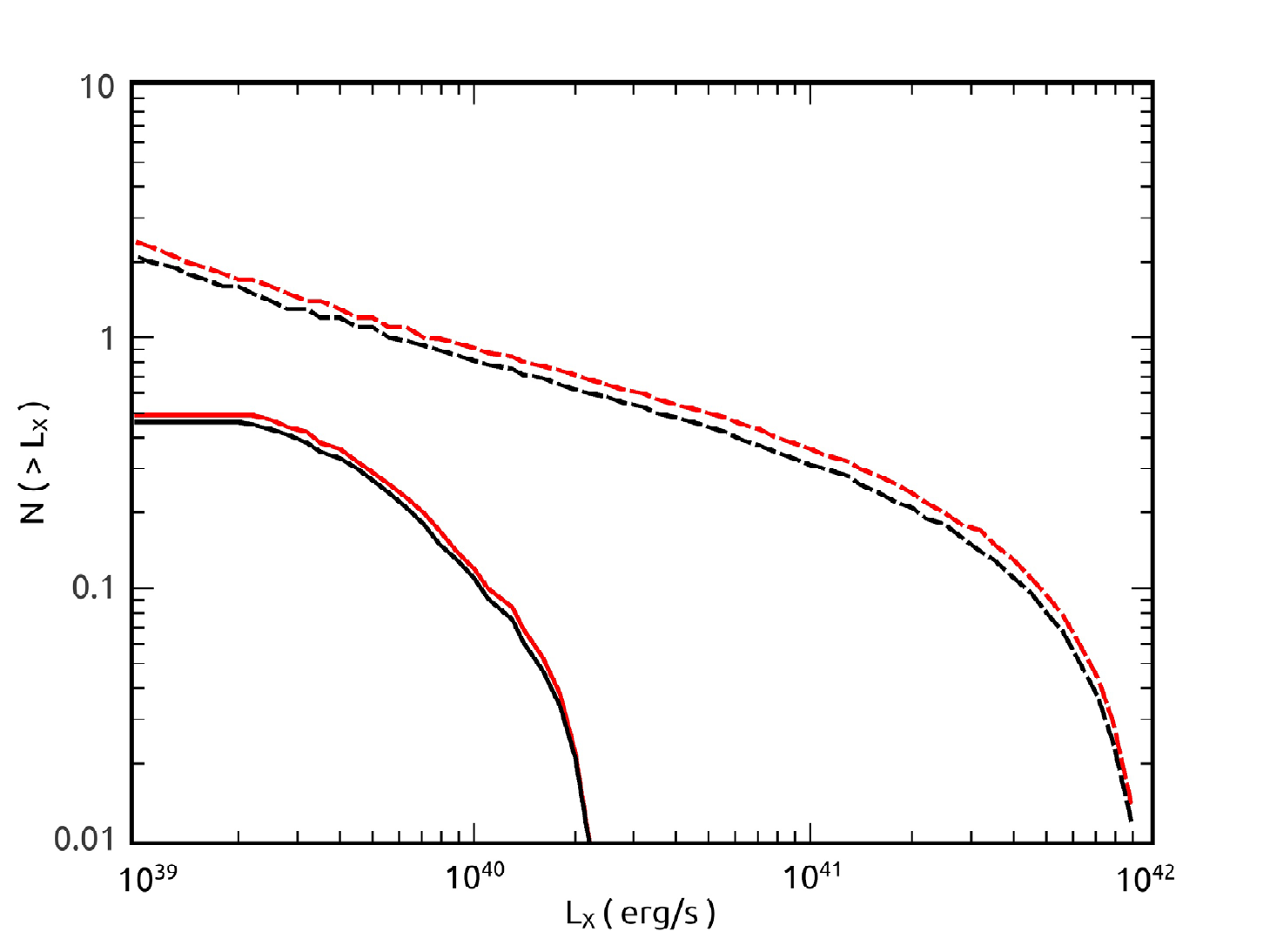}{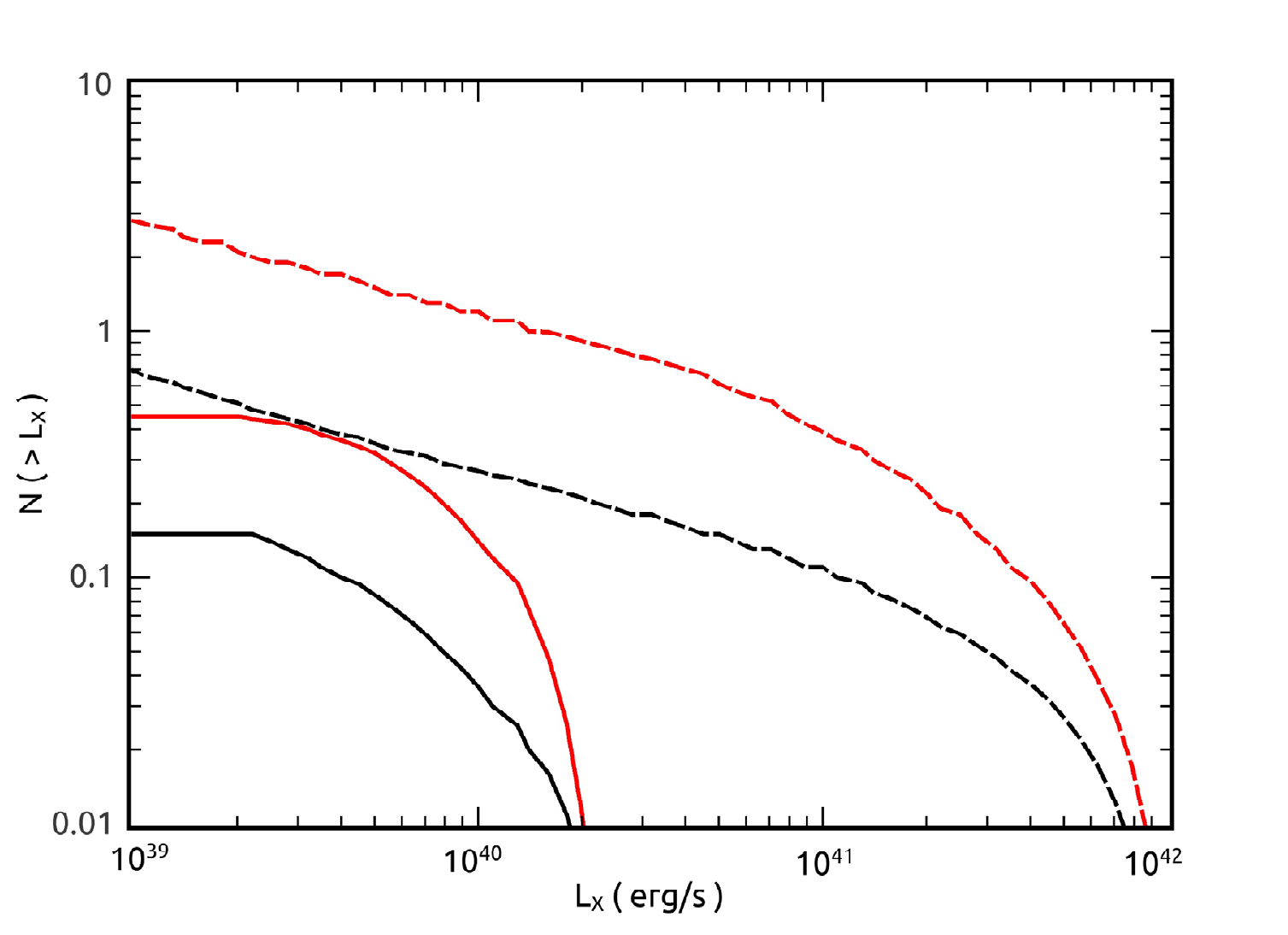}

\caption{The left and right panels show the X-ray luminosity
functions of the NS ULX population in case (1) and (2),
respectively. The red and black curves correspond to the donor mass
higher than $2 M_{\odot} $ and $5 M_{\odot} $, respectively. the
X-ray luminosities for the dashed and solid curves are calculated
with Eqs.~(2) and (4), respectively.\label{figure5}}

\end{figure}

\end{document}